\documentclass[seceq]{ptptex}

\usepackage{graphicx}
\usepackage{wrapft}

\newcommand{\beq}{\begin{eqnarray}}
\newcommand{\eeq}{\end{eqnarray}}
\newcommand{\del}{\partial}

\newcommand{\ket}{\rangle}
\newcommand{\bra}{\langle}

\newcommand{\Slash}[1]{\ooalign{\hfil/\hfil\crcr$#1$}}

\title{Structure of the nuclear force in a chiral 
quark-diquark model}

\author{%       %Use \scshape  for the family name
K. \textsc{Nagata}~\footnote{nagata@rcnp.osaka-u.ac.jp} and A. \textsc{Hosaka}
}

\inst{%         %Affiliation, neglected when [addenda] or [errata]
Research Center for Nuclear Physics (RCNP),
Osaka University, Ibaraki 567-0047, Japan
}

\abst{%         %this abstract is neglected when [addenda] or [errata]
We discuss the structure of the nuclear force using a   
lagrangian derived from hadronization of a chiral quark and diquark model.  
A generalized trace log formula including meson and nucleon fields 
is expanded 
to the order in which relevant terms emerge.  
It is shown that the nuclear force is composed of long and medium range 
parts of chiral meson exchanges and short range parts of quark-diquark 
exchanges. 
The ranges of the scalar and vector interactions coincide well with 
those of sigma ($\sigma$) and omega ($\omega$) meson exchanges 
if the size of the nucleon core of a quark-diquark bound state is  
adjusted appropriately.  
}

\begin{document}

\maketitle

%==================================================================
\section{Introduction}
%==================================================================

A microscopic description for the meson-nucleon interaction is 
one of central issues in the hadron-nuclear physics.  
Although lattice calculations seem to reach the level of 
quantitative description of hadron properties, 
interpretation based on effective models is still important 
in order to understand underlying physics mechanism.  
Symmetry is primarily a powerful tool when one attempts 
to understand physical meaning of some processes.  
In hadron physics, flavor and chiral symmetries are 
the two important symmetries which one should incorporate.  
In particular, inclusion of the pion degrees of freedom 
associated with spontaneous breaking of chiral symmetry is 
a crucially important ingredient.  
In addition, for baryon dynamics, their internal structure is also
important, as various form factors imply.  
These aspects can be intuitively understood by the fact that
pion Compton wavelength is larger than nucleon size and 
that nucleon size is larger than nucleon Compton wavelength:
$1/m_\pi >> \bra r_N^2\ket^{1/2} >> 1/M_N$.  

In a recent publication one method was proposed, 
which can deal with the above two aspects of chiral symmetry and 
internal structure of hadrons using an extended model of the 
Nambu-Jona-Lasinio model~\cite{NJL} including not only the quark-antiquark 
but also quark-diquark correlations~\cite{Abu-Raddad:2002pw}.  
It was shown that the method works well 
for both mesons and baryons.  
A microscopic lagrangian was path-integrated to generate 
an effective lagrangian for mesons and baryons with maintaining 
important symmetries such as the gauge and chiral symmetries.  
Hadron structure was then described by their constituents; 
a quark and an antiquark for mesons and a quark and a 
diquark for baryons.  
It was also pointed out that the resulting effective lagrangian 
contains various interactions among hadrons, such as meson-meson, 
meson-baryon and baryon-baryon interactions.  

In this paper, we investigate, among many interactions, 
nucleon-nucleon (NN) 
interaction\footnote{In what follows we 
use ``nucleon" rather than ``baryon", as we consider flavor SU(2) 
for the 
nucleon sector.} derived from the previous 
framework~\cite{Abu-Raddad:2002pw}.  
The NN interaction below the meson production threshold 
is phenomenologically well known as derived from the phase shift 
analysis, but its microscopic derivation is still needed.  
While the long range part is well described by meson exchange 
picture, there are several different approaches for the description of 
the short range part, as including meson exchanges and quark 
exchanges~\cite{Lacombe:dr,Machleidt:hj,Toki:ai,Oka:rj,Takeuchi:yz,Fujiwara:2001pw}.  
In our model, all components of the NN force is contained in 
the effective lagrangian which is written in a concise form 
of trace-log type.  
Then we will see that the NN interaction is composed of the long and 
medium range parts as described by meson exchanges, 
and short range part of quark-diquark exchanges.  
The latter is then shown to contain various components of 
the NN interaction such as central, spin-orbit and tensor types 
with non-locality.  

As explained in the next section, there are two types of diquarks; 
one is scalar, isoscalar diquark and the other is the axial-vector, 
isovector one.  
In the present paper, we perform analytic calculations including the  
both diquarks and discuss general structure of the NN interaction.  
However,  
for numerical computation, we consider only scalar diquark due to 
rather complicated structure when the axial-vector diquark is included.  
We then study interaction ranges in more detail than interaction strengths, 
since we expect 
that the former would be less sensitive to the kind of the diquark 
included. 

We organize this paper as follows.  
In section 2, we briefly show the derivation of the trace-log 
formula in the path-integral hadronization of the NJL model
including quark-diquark correlations.  
In section 3, terms containing the NN interaction are investigated 
in detail, 
where general structure of the NN amplitude is presented.  
We show a sample numerical calculation only for the case 
containing a scalar diquark.  
The present study of the NN interaction is not quantitatively 
complete but will be 
useful in showing some important aspects of the nuclear force, 
in particular that the range of the short range interaction is 
related to the intrinsic size of the nucleon.  
The final section is devoted to summary.

%==================================================================
\section{Effective lagrangian for mesons and nucleons}
%==================================================================

We briefly review the method to derive 
an effective lagrangian for mesons and nucleons  
from a quark and diquark model of chiral symmetry following the 
previous work by Abu-Raddad et al~\cite{Abu-Raddad:2002pw}.
Let us start from the NJL Lagrangian
%%%
\begin{equation}
{\cal L_{\text{NJL}}} = {\bar q} (i\rlap/\partial -m_0) q + \frac{G}{2} \left[
({\bar q} q)^2 + ({\bar q} i \gamma_5 \vec{\tau} q)^2 \right]\;.
\end{equation}
%%%
Here $q$ is the current quark field, $\vec{\tau}$  the isospin
(flavor) Pauli matrices, $G$
a dimensional coupling constant, and $m_0$ the  
current quark mass. 
In this paper we set $m_0 = 0$ for simplicity.  
As usual, the NJL lagrangian is bosonized 
by introducing collective meson fields as auxiliary fields in the 
path-integral method~\cite{Eguchi:1976iz,Dhar:1983fr,Ebert:1985kz}.  
At an intermediate step, we find the following lagrangian:  
\beq
{{\cal L}^\prime_{q\sigma\pi}} = 
\bar{q}\left( i\rlap/\partial  
- (\sigma + i\gamma_5 \vec\tau\cdot\vec\pi) \right) q
-\frac{1}{2G}(\sigma^2+\vec\pi^{\; 2})\, . 
\label{Lprime}
\eeq
Here $\sigma$ and $\vec \pi$ are 
scalar-isoscalar sigma
and pseudoscalar-isovector pion fields as generated from 
$\sigma \sim \bar q q$ and $\vec \pi \sim i\bar q \vec \tau \gamma_5q$, 
respectively.  
For our purpose, 
it is convenient to work in the non-linear basis rather than 
linear one~\cite{Ebert:1997hr,Ishii:2000zy}.
It is achieved by the chiral rotation from the current ($q$) to 
constituent ($\chi$) quark fields: 
\beq
\chi = \xi_5 q\, , \; \; \; \; 
\xi_5 
= 
\left(
\frac{\sigma + i \gamma_5 \vec \tau \cdot \vec \pi}{f} 
\right)^{1/2} \, , 
\eeq
where $f^2 = \sigma^2 + \vec \pi^{\; 2}$.  
Thus we find 
\beq
{\cal L^\prime_{\chi \sigma \pi}} = 
\bar{\chi}\left(i\rlap/\partial  
- f - \rlap/v - \rlap/a\right) \chi
-\frac{1}{2G} f^2 \, , 
\label{Lprime2}
\eeq
where 
\beq
v_\mu = - \frac{i}{2} \left(
\del_\mu \xi^\dagger \xi +  \del_\mu \xi \xi^\dagger 
\right)\, , \; \; \; 
a_\mu = - \frac{i}{2} \left(
\del_\mu \xi^\dagger \xi -  \del_\mu \xi \xi^\dagger 
\right)\, , 
\eeq
are the vector and axial-vector currents written in terms of the 
chiral field
\beq
\xi_5 
= 
\left(
\frac{\sigma + i \vec \tau \cdot \vec \pi}{f} 
\right)^{1/2}
\eeq
The lagrangian (\ref{Lprime2}) describes not only the free
kinetic term of the quark, but also quark-meson interactions 
such as Yukawa type, Weinberg-Tomozawa type and etc.  

In a model we consider here, we introduce 
diquarks and their interaction terms with quarks.  
We assume local interactions between quark-diquark
pairs to generate the nucleon field.  
As inspired from the method to construct local nucleon 
fields, it is sufficient to consider two diquarks; 
one is scalar, isoscalar one, $D$,  and the other is an 
axial-vector, isovector one, $\vec D_\mu$~\cite{espriu}.  
Hence, our microscopic lagrangian for quarks, diquarks and 
mesons is given by 
\begin{eqnarray}
{\cal L} &=& \bar{\chi}(i\rlap/\del - f 
- \rlap/v - \rlap/a) \chi \;-\;
\frac{1}{2G}f^2\;+
D^\dag (\del^2 + M_S^2)D \;+\; 
{\vec{D}^{\dag\;\mu}} 
\left[  (\del^2 + M_A^2)g_{\mu \nu} - \del_\mu \del_\nu\right]
\vec{D}^{\nu} \nonumber\\ &&
\;+\; \tilde{G} \left( \sin{\theta} \; \bar{\chi}\gamma^\mu \gamma^5
\:\vec{\tau}\cdot {\vec D}^\dag_{\mu} \;+\; \cos{\theta}
\;\bar{\chi} D^\dag\right)\; \left(\sin{\theta}\; {\vec D}_{\nu}\cdot
\vec{\tau} \: \gamma^\nu \gamma^5 \chi \;+\; \cos{\theta} \;
D \chi \right)\; .
\label{lsemibos}
\end{eqnarray}
In the last term $\tilde G$ is another coupling constant for the 
quark-diquark interaction and the angle $\theta$ 
controls the mixing ratio of the scalar and axial-vector diquarks
in the nucleon wave function.   

One would have constructed a nucleon field in the linear basis.  
However, this makes chiral transformation properties 
complicated.  
In fact, if we replace $\chi$ by $q$, then more than two 
terms are necessary to maintain chiral symmetry for the 
interaction term in (\ref{lsemibos}). 
In contrast, in the non-linear basis, the transformation 
property becomes simple; both quark and nucleon fields 
are isospin multiplets which are subject to 
the non-linear transformation
\beq
\chi(x) \to h(x) \chi(x) \, , \; \; \; B(x) \to h(x) B(x) \, .
\eeq
Here $h$ is a non-linear function of the chiral 
transformations and the chiral field at a point 
$x$~\cite{weinberg_book,hosaka_book}.  

Now the hadronization procedure, the elimination of the 
quark and diquark fields in (\ref{lsemibos}), is straightforward.  
The final result is written in a compact form as~\cite{Abu-Raddad:2002pw}
%%%
\begin{eqnarray}
{\cal L}_{\rm eff} &=& 
- \frac{1}{2G}f^2 
\;-\; i \;{\rm tr\; ln} (i \rlap/\del - f -\rlap/v - \rlap/a)
\;-\; \frac{1}{\tilde{G}}\; \bar{B} B \;+\; 
i\;{\rm tr \; ln} ( 1\;-\; \Box ) \, .
\label{effL}
\end{eqnarray}
%%%
Here trace is over space-time, color, flavor, and Lorentz indices, and 
the operator $\Box$ is defined by 
\beq
\label{eqbox}
\Box & = & 
\begin{pmatrix} {\cal A}& {{\cal F}_2}\\
 {{\cal F}_1} &{\cal S}
\end{pmatrix} \, ,
\eeq
where
\begin{subequations}
\begin{eqnarray}
{\cal A}^{\mu i,\,\nu j} & = & {\sin}^2\theta \
\bar{B} \; 
\gamma_\rho \gamma^5\;
  {\tau}_{k} \;\tilde{\Delta}^{\rho k,\,\mu i} \;
 S\; {\tau}^{j}\;
  \gamma^\nu \gamma^5\; B\;,\\ 
{\cal S}& = & {\cos}^2\theta\; \bar{B}\;\Delta
  \;S\;B\;,\;\; \;\;\;\;\\ 
({{\cal F}}_1)^{\nu j}& = & \sin{\theta} \cos{\theta}\;\bar{B}\; 
\Delta \;S\; 
 {\tau^j}\;\gamma^\nu \gamma^5\;  B\;,\\
({\cal F}_2)^{\mu i}& = & \sin{\theta} \cos{\theta} \;
\;\bar{B} \;\tilde\Delta^{\rho k,\, \mu i}\;\gamma_\rho \gamma^5\; {\tau_k} \; S\; B\;.
\end{eqnarray}
\end{subequations}

\begin{figure}
       \centerline{\includegraphics[width=5cm]
                                   {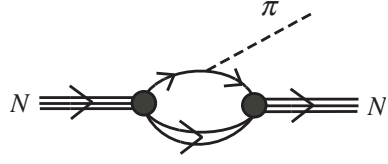}}
\centering
\begin{minipage}{14cm}
   \caption{\small 
   Meson-nucleon Yukawa vertex. }
   \label{mbb}
 \end{minipage}
\end{figure}

\begin{figure}
       \centerline{\includegraphics[width=8cm]
                                   {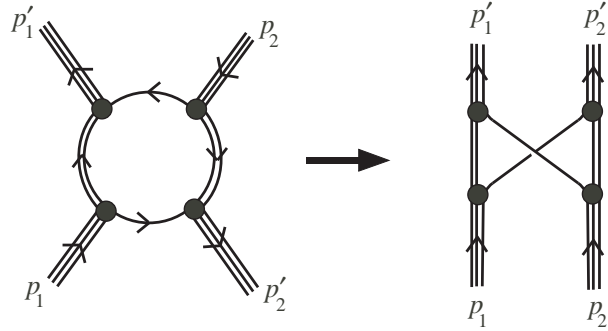}}
\centering
\begin{minipage}{14cm}
   \caption{\small 
   A loop diagram for the NN interaction (left) and the equivalent 
quark exchange diagram.  }
   \label{bbbbloop}
 \end{minipage}
\end{figure}

Though the effective meson-nucleon lagrangian (\ref{effL}) 
looks simple, 
it contains many important physics ingredients 
when the trace-log
terms are expanded:
\begin{itemize}
\item
It generates a meson
lagrangian in a chirally symmetric manner.  
Up to the fourth power in the meson fields, it produces precisely 
the lagrangian of the linear sigma model with realization of 
spontaneous breaking of chiral symmetry.

\item
From the second trace-log term, a nucleon effective lagrangian 
is derived.  
In the previous paper, kinetic term of the nucleon was 
investigated and nucleon mass was computed at the one-loop 
level~\cite{Abu-Raddad:2002pw}.

\item
In the nucleon effective lagrangian, meson-nucleon couplings 
appear through the diagrams as shown in Fig.~\ref{mbb}.  
Their strengths and form factors are computed 
by the underlying quark-diquark dynamics.  
Using these vertices, meson-exchange interactions 
are constructed.  

\item
There are diagrams which contain many nucleon fields.  
For instance, NN interactions are expressed 
by one-loop diagram as shown in Fig.~\ref{bbbbloop}.  
This term describes a short range part of the NN 
interaction.  
In this paper, we focus our attention mostly on the NN 
interaction derived from the one-loop diagrams.  

\end{itemize}

%figure1-------Yukawa-vertex------------------------
%figure---------------------------------------------

%figure2-------box diagram---------------------------
%figure---------------------------------------------

%==================================================================
\section{The structure of the NN interaction}
%==================================================================

As we have explained, the NN interaction in the 
quark-diquark model consists of two parts: 
one is the meson exchange terms for long and medium range interactions 
and the other is the one described 
by the quark-diquark one-loop diagram shown in Fig.~\ref{bbbbloop} for 
the short range interactions.  
The latter diagram is equivalent, when the flow of nucleon lines 
are arranged appropriately, to quark or diquark exchange diagrams
as shown in Fig.~\ref{bbbbloop}.
In this quark exchange process, the nucleon goes into the 
quark-diquark pair through the interaction (blobs in the diagram).  
In this sense, this process differs from the norm kernel of 
quark exchange in a quark cluster model~\cite{Toki:ai,Oka:rj}.

%---------------------------------------
\subsection{One pion exchange interaction}
%---------------------------------------

Let us first examine the long rage part which is dominated by the 
one pion exchange.  
In the present model setup, we start from the Nambu-Jona-Lasinio
model which includes the pion and sigma mesons.   
In the nuclear force, the scalar-isoscalar interaction  
also emerges from quark exchanges at the same energy scale as
the sigma meson exchanges. 
This is a general remark when we consider the medium and short 
range parts of the nuclear force; there are two contributions 
from boson exchanges and quark exchanges, a natural consequence 
when nucleon consists of quark and diquark core  
surrounded by meson clouds.  

Now the pion exchange process in the present model is 
supplemented by the form factor due to the quark-diquark structure 
of the nucleon as shown in Fig.~\ref{mbb}.  
In the momentum space, the one pion exchange interaction
is then written as, in the non-relativistic approximation, 
\beq
V_{\rm OPEP}(\vec q) = 
\frac{g_{\pi NN}(\vec q^{\; 2})}{2M_N} 
\frac{\vec \sigma_1 \cdot \vec q \; \vec \sigma_2 \cdot \vec q}
{\vec q^{\; 2} + m_\pi^2} 
\frac{g_{\pi NN}(\vec q^{\; 2})}{2M_N} \, , 
\eeq
where 
$g_{\pi NN}(\vec q^{\; 2})$ is the $\pi NN$ form factor, 
$M_N$ the mass of the nucleon and 
$\vec \sigma_{1, 2}$ are the spin matrices for the nucleon 1 and 2.    
Due to chiral symmetry, 
it is related to the axial form factor through the Goldberger-Treiman 
relation~\cite{Goldberger:1958tr}
\beq
g_{\pi NN} = \frac{M_N}{f_\pi} g_A \, .
\eeq

In the present paper, we include only the scalar diquark 
due to rather complicated structure of the quark-diquark loop integral 
when the axial-vector diquark is included.  
A full calculation with both scalar and axial-vector diquarks included 
requires careful treatment of the regularization 
procedure.  
We will postpone this complete study in future works.  
It is, however, not very difficult to estimate the effect 
of the axial-vector diquark for one-body matrix elements; 
it mostly contributes to the absolute values of spin-isospin 
($\vec \sigma \cdot \vec \tau$) matrix elements, which are axial-vector 
coupling constant $g_A$ (and hence the $\pi NN$ coupling constant 
$g_{\pi NN}$) and isovector magnetic moments.  
For quantities which reflect the size of the quark-diquark 
bound state such as $\vec q^{\; 2}$ dependence of form factors are not 
very sensitive to the kind of diquarks, since they are dominated 
by the extension of the lighter particle, the quark.  
%For two body matrix elements such as the NN interaction, 
%once again the quantities related to the size of the quark-diquark 
%distribution such as the 
%However, apart from absolute strengths of the interactions, we
%expect that the range of the matrix element
%can be studied in a more stable manner even though 
%only the scalar diquark is considered; 
%both scalar and axial-vector diquarks form a nucleon core of a typical size 
%about 0.5 $\sim$ 0.6 fm.   

To start with, numerical calculations are performed using 
the parameters as shown in Table \ref{tab:parameter}.  
These were used in the previous work for the study of a
single nucleon~\cite{Abu-Raddad:2002pw}.  
The divergent integrals are regularized by the Pauli-Villars
method with the cut-off mass $\Lambda_{PV}$~\cite{izykson}.  
Then we perform calculations 
by varying the coupling constant for the quark-diquark 
interaction $\tilde G$.
The strength of $\tilde G$ controls the binding or size of 
the nucleon as shown in Table 1.  
Here the size is defined as a sum of the quark and diquark 
distributions:
\beq
\bra r^2 \ket = \bra r^2_q\ket + \bra r^2_D\ket.  
\eeq
Note that this is different from the charge distribution 
where each term is weighted by the corresponding charge.  
We will see how physical quantities such as 
form factors and short range interactions are related to the 
extension of the quark-diquark wave functions.  

%table 1-------Laith's parameters--------------------------
\begin{table}[tbh]
{\small
\begin{center}
\begin{minipage}{14cm}
\caption{
Model parameters taken from Ref. \cite{Abu-Raddad:2002pw}\label{tab:parameter}}
\end{minipage}
\begin{tabular}{ccccc}
\vspace*{-3mm}\\
\noalign{\hrule height 0.8pt}
$M_N$& $m_q$& $M_D$& $\Lambda_{PV}$& $\tilde{G}$\\
\hline
0.94 GeV & 0.39 GeV & 0.60 GeV & 0.63 GeV &271.0 GeV$^{-1}$\\
\noalign{\hrule height 0.8pt}
\end{tabular}
\end{center}
}
\end{table}
%table--------------------------------------------------

%Table 2. parameters used in nn force-----------------
\begin{table}[tbh]
{\small
\begin{center}
\begin{minipage}{14cm}
\caption{Basic nucleon properties when $\tilde G$ is varied. 
Other model parameters $m_q$, $M_D$ and $\Lambda$ are fixed.
\label{tab:bind_vs_couple}}
\end{minipage}
\begin{tabular}{cccc}
\vspace*{-3mm}\\
\noalign{\hrule height 0.8pt}
$\tilde{G}$(GeV$^{-1}$)& B.E.(MeV) & $M_N$(MeV)& $\langle r^2\rangle^{1/2}$(fm)\\
\hline
156.4& 10 & 980& 1.64\\
271.0& 50 & 940& 0.89\\
445.9& 140& 850& 0.60\\
529.4& 190& 800& 0.54\\
609.2& 240& 750& 0.51\\
687.4& 290& 700& 0.48\\
\noalign{\hrule height 0.8pt}
\end{tabular}
\end{center}
}
\end{table}
%table------------------------------------------------------------

\begin{figure}
       \centerline{\includegraphics[width=7cm]
                                   {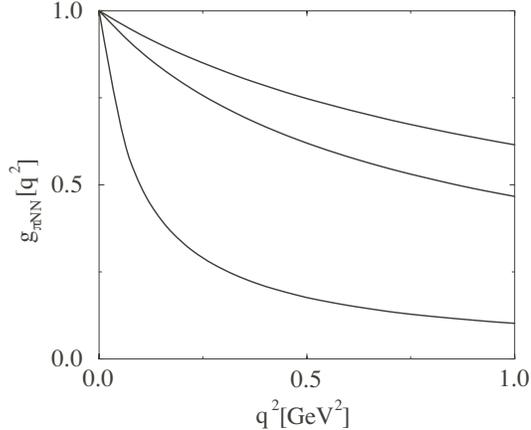}}
\centering
\begin{minipage}{14cm}
   \caption{\small 
   Pion-nucleon form factor $g_{\pi NN}$. 
   Three lines are for $M_N$=0.98 (loosely bound), 0.85, 0.70 (tightly bound)
   GeV from bottom to top.}
   \label{piNNff}
 \end{minipage}
\end{figure}

In Fig.~\ref{piNNff}, we plot the $\pi NN$ form factor
as a function of the square momentum transfer $\vec q^{\; 2}$
for different binding energies. 
We did not attempt to reproduce the experimental value of 
the nucleon mass, since our interest here is how the form factor 
is related to the extension of the nucleon wave function due to 
the quark-diquark structure.  
The result is the same as the previous work when 
the parameters shown in Table~\ref{tab:parameter} are used.
As $\tilde G$ is increased such that the size of the nucleon
is reduced, the $q$-dependence of the form factor
becomes weak.
We have extracted the cut-off parameter of the form
factor when fitted by a monopole function,  
$\Lambda^2/(\Lambda^2 + \vec q^{\; 2})$, and the range of the 
axial form factor, $\bra r^2\ket^{1/2}_A$.  
The result is shown in the second and third 
column of Table~\ref{tab:axialrange}.  
When the previous parameter set is used such that the nucleon 
(electromagnetic) size about 0.8 fm is reproduced  
by the quark-diquark wave function~\cite{Abu-Raddad:2002pw}, 
the form factor becomes
rather soft, $\Lambda \sim 500$ MeV.  
However, if we consider the fact that meson clouds should also 
contributes to the nucleon size and the core part should be
smaller than the observed size, then the quark-diquark 
extension could be smaller.  
If we take, for instance, the mean square value of 
the quark-diquark extension about 
$(0.6 \; {\rm fm})^2$, then the 
cut-off size of about 800 GeV is obtained. 
This value is smaller as compared with those often used in the 
NN force~\cite{Machleidt:1987hj}.  
However, the cut-off parameters in the NN force depends 
on the modeling of the meson exchanges.  
In fact, the above value is consistent with that
extracted from the electroproduction of 
the pion~\cite{Liesenfeld:1999mv}.

%%Table 3. Axial cutoff ------------------------------------------
\begin{table}[tbh]
\begin{center}
\caption{{\footnotesize The cut-off parameter and the corresponding range
of the $\pi NN$  form factor.}\label{tab:axialrange}}
\begin{tabular}{ccc}
\vspace*{-3mm}\\
\noalign{\hrule height 0.8pt}
$M_N$(MeV)& $\Lambda$(GeV)&$\bra r^2\ket^{1/2}_A$(fm$^{-1}$)\\
\hline
 980& 0.29& 1.67\\
 940& 0.54& 0.90\\
 850& 0.78& 0.62\\
 800& 0.85& 0.57\\
 750& 0.90& 0.53\\
 700& 0.93& 0.52\\
\noalign{\hrule height 0.8pt}
\end{tabular}
\end{center}
\end{table}
%%table------------------------------------------------------------

%figure3-----NNpi form factor------------------------
%figure---------------------------------------------

%---------------------------------------
\subsection{Short range interaction}
%---------------------------------------

Let us turn to the short range interaction described by the 
quark (or diquark) exchanges, as shown in Fig.~\ref{bbbbloop}.   
Using the interaction vertices given in Eq.~(\ref{eqbox}), 
it is straightforward to compute the amplitude for the 
quark-diquark loop:
\begin{eqnarray}
{\cal M}_{NN} &=& -i N_c Z^2 \nonumber \\
& & \hspace*{-2cm} \times \int
\frac{d^4k}{(2\pi)^4}\frac{\bar{B}(\Slash{k}+m_q)B\bar{B}(\Slash{p}_2-\Slash{p}_2^{\prime}+\Slash{k}+m_q)B}
{[(p_1-k)^2-M_s^2][k^2-m_q^2][(p_2^\prime -k)^2-M_s^2][(p_2-p_2^{\prime}+k)^2-m_q^2]}\ ,
\label{eqn:NNN}
\end{eqnarray}
where $N_c$ is the number of colors and $Z$ is the wave function 
normalization of the nucleon field~\cite{Abu-Raddad:2002pw}.  
In this paper, we evaluate the integral in the center-of-mass system
for the elastic scattering of the two nucleon:
\begin{eqnarray*}
p_1&=&(E_{\vec{p}},\ \vec{p}),\ p_2=(E_{\vec{p}},\ -\vec{p})\\
p_1^{\prime}&=&(E_{\vec{p}^{\; \prime}},\ \vec{p}^{\; \prime}),\ 
p_2^{\prime}=(E_{\vec{p}^{\; \prime}},\
-\vec{p}^{\; \prime})\; , \; \; \; 
|\vec p| = |\vec p^{\; \prime}| \; .  
\end{eqnarray*}
In principle, from this amplitude we should calculate observables such as 
phaseshifts, cross sections for physical two-nucleon channels by properly 
taking into account anti-symmetrization.  
Since in the present calculation we include only the scalar diquark, we do 
not expect that the comparison of such quantities with data will make 
significant sense.  
Rather, in the following we study some basic properties of the amplitude
itself, mostly the interaction ranges extracted from  Eq.~(\ref{eqn:NNN}).  

To do so, we write 
the amplitude (\ref{eqn:NNN}) in a schematic manner as
\beq
{\cal M}_{NN}
&=&
F_S(\vec{P},\vec{q})(\bar{B}B)^2+F_V(\vec{P},\vec{q})
(\bar{B}\gamma_{\mu}B)^2+\cdots\
\label{eqn:NNforce} ,
\eeq
where
\beq
\label{defPq}
\vec{P}=\vec{p}^{\; \prime}+\vec{p}
\, \; \; \; 
\vec{q}=\vec{p}^{\; \prime}-\vec{p}
\eeq
In Eq.~(\ref{eqn:NNforce}), we identify $F_S$ and $F_V$ with the 
scalar and vector interactions, respectively.  
Dots of Eq.~(\ref{eqn:NNforce}) then 
contain terms involving external momenta $p_i$ such as 
$\bar B \Gamma(p_i) B$, where  $\Gamma(p_i)$ is a $4 \times 4$ matrix 
involving $p_i$.  
The latter are usually not considered in the local potential
but here they exist due to the non-local nature of the amplitude
(\ref{eqn:NNN}).  
Furthermore, we note that the bilinear forms of the nucleon fields
in Eq.~(\ref{eqn:NNforce}) are taken by the pair of $B$ fields 
in Fig.~\ref{bbbbloop} carrying 
the momentum $p_1$ and $p_2^\prime$, and the other 
carrying the momentum $p_2$ and $p_1^\prime$.  
This is a consequence of the quark-diquark loop integral.  
Before going to details, it is pointed out that 
the scalar interaction is attractive, while the vector interaction is 
repulsive. 
This is a general feature of the interaction between 
two nucleons (fermions) and is independent of the details of a model. 

\begin{figure}[tbh]
\centering{
\includegraphics[width=11.0cm]{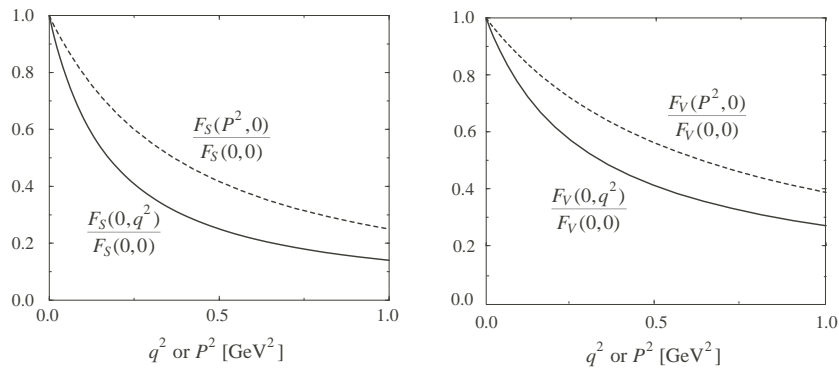}
}
\begin{minipage}{14cm}
\caption{\small 
Normalized scalar and vector form factors
as function of $q^2 = |\vec q^{\; 2}|$ with fixed 
$\vec P = 0$ (solid line) and 
those of $P^2 = |\vec P^{\; 2}|$ with fixed 
$\vec q = 0$ (dashed line). 
Calculations are performed using the parameter set of 
Table 1. }
\label{ffscalar}
\end{minipage}
\end{figure}

In general the amplitude is highly non-local as the symmetric 
loop diagram implies.  
The loop diagram may be interpreted as either a quark exchange or 
diquark exchange diagram.  
It is then natural that the heavier diquark exchange is 
of shorter range nature.  
By identifying the momentum transfer exchanged by a quark pair as 
${\vec q}$ and that of the diquark pair by ${\vec P}$,   
it is expected that ${\vec P}$ dependence is milder than 
${\vec q}$ dependence.  
We show this explicitly in Fig.~\ref{ffscalar}.  
Therefore, we expand the functions $F_{S,V}$ in powers of 
${\vec P}$ and evaluate their coefficients as functions of 
${\vec q}$:
\beq
\label{Vexpand}
F_i(\vec P, \vec q) =F_i(\vec P, \vec q)\vert_{\vec P = 0}
+ \vec P \cdot \frac{\del}{\del \vec P} 
F_i(\vec P, \vec q)\vert_{\vec P = 0} + \cdots \, ,
\eeq
where $i$ stands for $S$ or $V$.  
The resulting ${\vec q}$ dependent functions might be interpreted as 
a Fourier transform of a local potential as a function of the relative 
coordinate ${\vec r} = {\vec x}_1 - {\vec x}_2$.  
%This outlines how an approximate local NN interaction is obtained.  
Then the general structure of the 
NN potential can be studied by 
performing non-relativistic reduction of 
the amplitude, Eqs.~(\ref{eqn:NNforce}) or (\ref{Vexpand}).  
It contains central, spin-orbit, tensor components as well as 
non-local terms as proportional to $\vec P$.   

%figure4---scalar and vector form factors--------------
%---------quark ex vs diquark ex----------------------
%figure---------------------------------------------

%After the non-relativistic reduction of 
%, we find
%\beq
%- {\cal L} \sim 
%V(\vec q) = (central) + (ls) + (tensor) + (energy dependent)
%\eeq
%As expected, the potential derived from the NN amplitude contains
%essentially all the terms required from the local-equivalent 
%potential.  
%Hence in principle, it is possible to predict all the components in the 
%nuclear force systematically from the loop diagrams.  

Now we discuss the scalar-isoscalar functions 
$F_S$ and $F_V$ in Eq.~(\ref{eqn:NNforce}).
We can not write the resulting $q$ dependence in a closed form,
but have to resort to numerical computations.
We have performed the Feynman integral numerically for 
several different $\tilde G$ parameters which yields different binding 
energies, or the size of the quark-diquark distribution.  
Although the one-loop integral of Fig.~\ref{bbbbloop} converges 
for a scalar diquark, we keep the counter term of the 
Pauli-Villars regularization.  
Since the present theory is a cut-off theory with 
the relatively small cut-off mass $\Lambda_{PV} = 0.63$ GeV, 
the counter term plays a significant role in order to produce 
numbers.

\begin{figure}[tbh]
\centering{
\includegraphics[width=5cm]{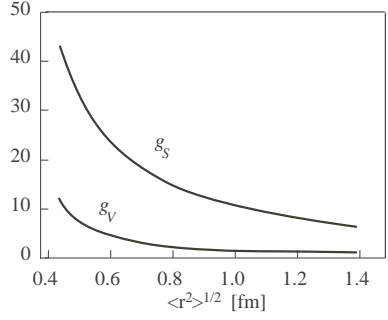}
}
\begin{minipage}{14cm}
\caption{\small 
Scalar and vector coupling constants as functions of the 
nucleon size $\bra r^2 \ket^{1/2}$. }
\label{gSV}
\end{minipage}
\end{figure}

First we see the strength of the interaction, by extracting 
the Yukawa coupling constant of the vector and scalar type
from the amplitude.  
The results are 
shown in Fig.~\ref{gSV} as functions of the 
size of the nucleon $\bra r^2 \ket$.  
These numbers may be compared with the empirical values
$g_S \sim 10$ and $g_V \sim 13$~\cite{Machleidt:1987hj}.  
The present results are strongly dependent on $\bra r^2 \ket$.  
When only scalar diquark is included the scalar interaction 
becomes much stronger that the strength of the vector interaction.  
Phenomenologically, the vector (omega meson) coupling is slightly 
stronger than the scalar (sigma meson) coupling.  
We expect that the present result will be changed significantly 
when axial-vector diquark is included.  

%figure-------strength vs <r2>1/2---------------
%figure---------------------------------------------

\begin{figure}[tbh]
\centering{
\includegraphics[width=5cm]{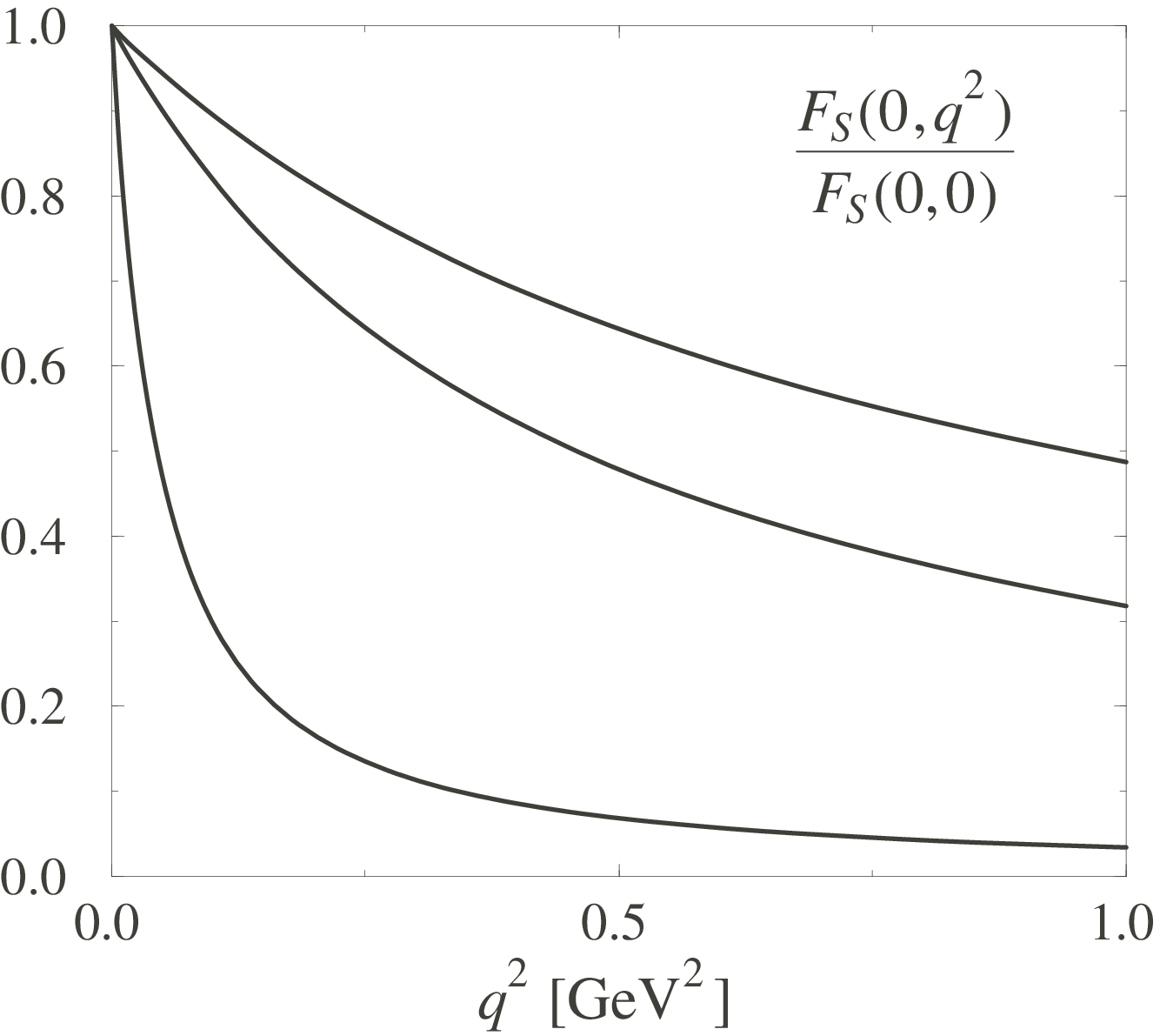}
\; \; \; 
\includegraphics[width=5cm]{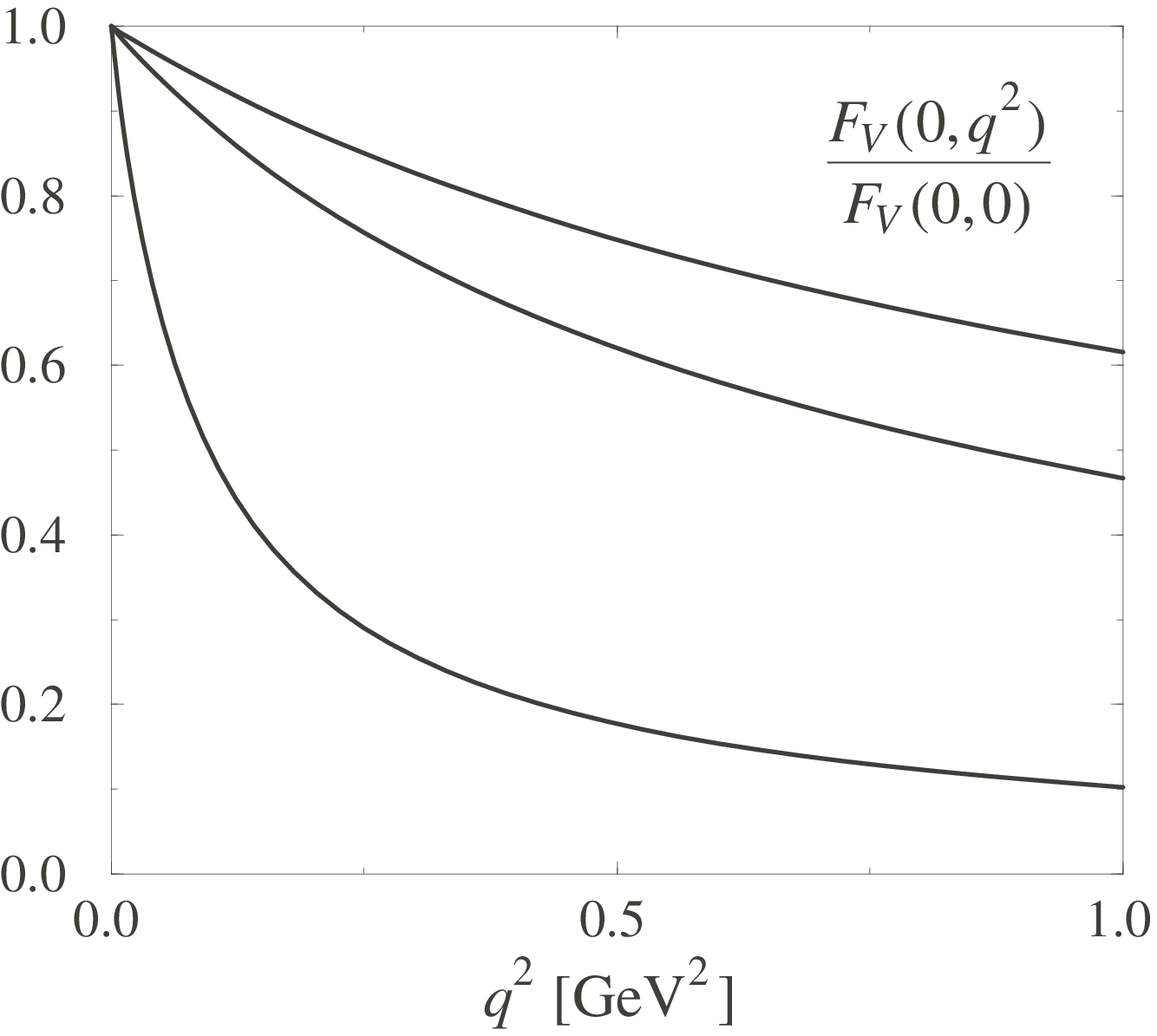}
}
\begin{minipage}{14cm}
\caption{\small 
Normalized scalar and vector form factors 
$F_{S,V}(q^2,\ P^2=0)/F_{S,V}(q^2=0,\
P^2=0)$. Each lines are for $M_N$=0.98, 0.85, 0.70 GeV from bottom 
to top.\label{fig5scalar}}
\end{minipage}
\end{figure}

\begin{figure}
\centering{
\includegraphics[width=5cm]{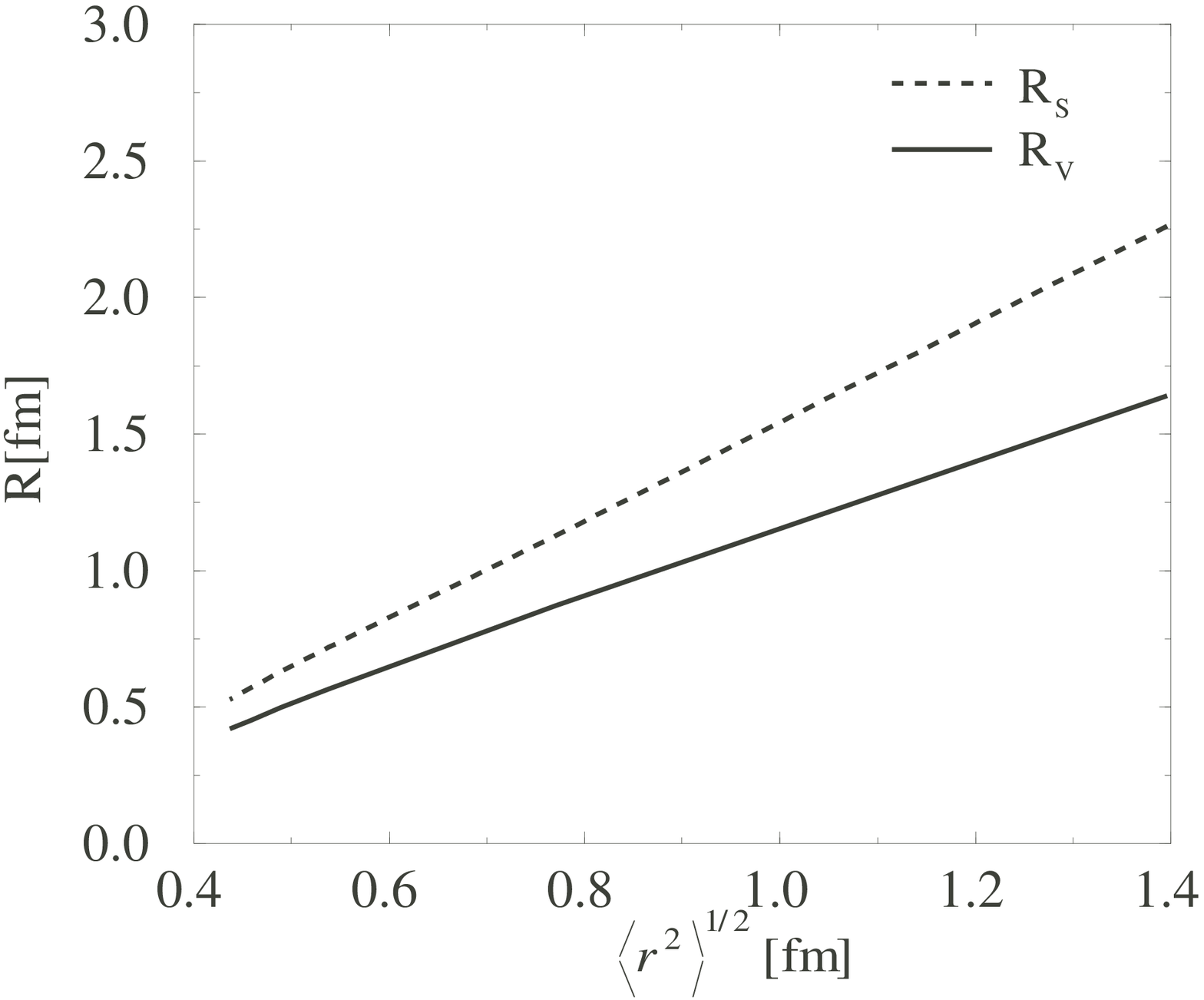}
\hspace*{1cm}
\includegraphics[width=5cm]{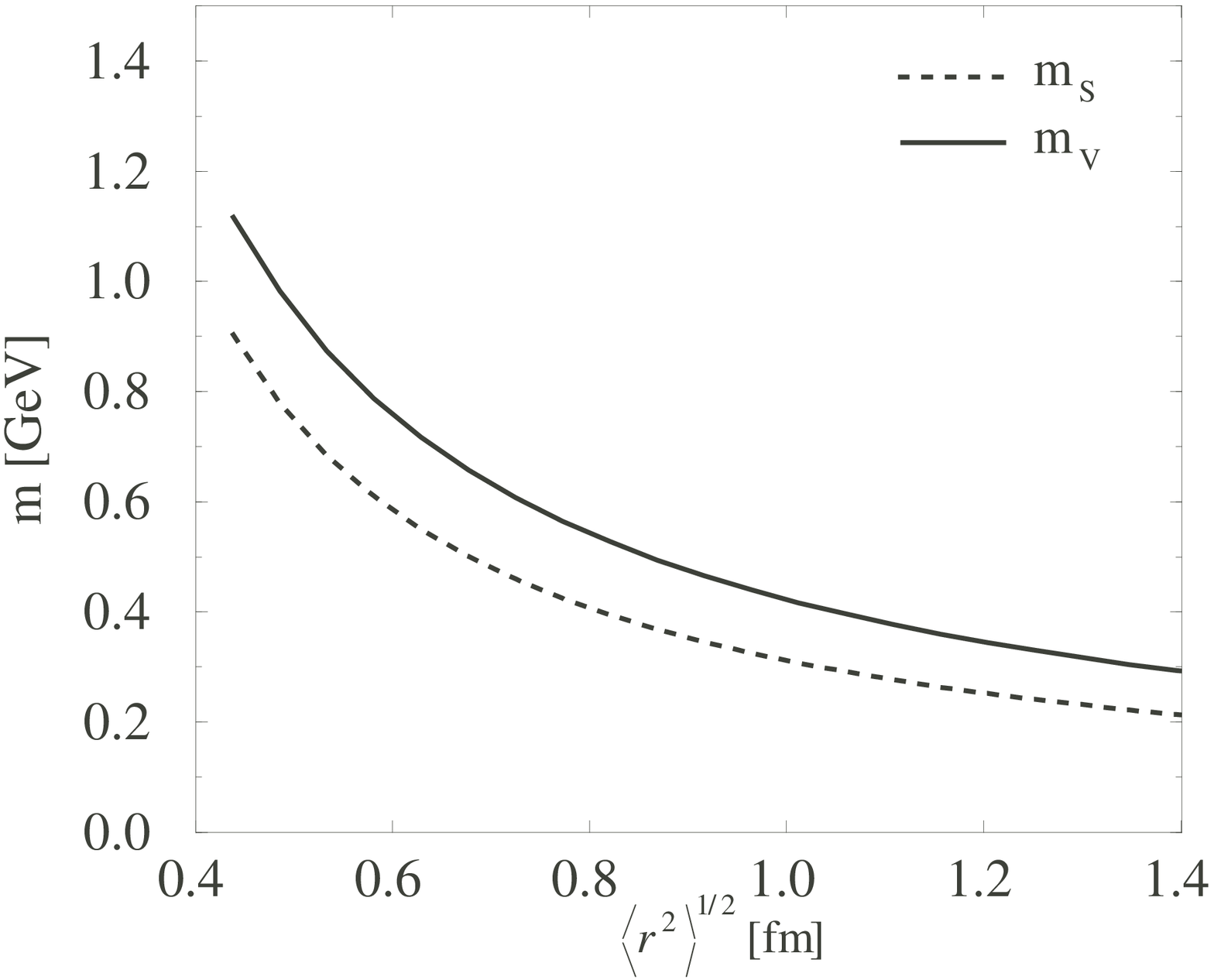}}
\begin{minipage}{14cm}
\caption{\small
 The interaction range $R_i$ (left panel) and the corresponding 
mass parameter $m_i$ (right panel) as functions of the nucleon size
(distribution of a quark and a diquark.)}
   \label{fig:sizevsrange}
\end{minipage}
\end{figure}

Let us turn to the discussion of the interaction ranges which is 
expected to be more reliably studied than the strength 
even when only the scalar diquark is included.   
In Fig.~\ref{fig5scalar}, we have shown the $q$-dependence 
of the zeroth order coefficients of (\ref{Vexpand}) for three 
different sizes of the nucleon.  
It is obvious from Fig.~\ref{fig5scalar}, 
as the size of the nucleon becomes smaller the interaction range 
becomes shorter range.   
More quantitatively, we define the interaction range $R_i$ by 
\beq
R_i^2 \equiv - 6 \left.\frac{1}{F_i(q^2)}\frac{\del F_i}{\del
q^2}\right|_{q^2\rightarrow 0} \, ,  
\label{defrange}
\eeq
which is related to the mass parameter of the interaction range by
$m_i \equiv \sqrt{6}/R_i$.  
%At the same time we define the size of the nucleon as described by the 
%quark-diquark bound state $\bra r^2 \ket$.  
%In Table~\ref{tab:sizevsrange} we show the result for $R_i$ 
%and $m_i$.
It is interesting to see that the mass (and hence range)
parameters of the interaction are approximately 
proportional to the size of the nucleon $\bra r^2\ket^{1/2}$
as shown in Fig.~\ref{fig:sizevsrange}.

%%table--------range,mass,strength and size---------
%\begin{table}[tbh]
%\begin{center}
%{\small
%\begin{minipage}{12cm}
%\caption{Interaction ranges and strengths as functions vs.
%the size of the nucleon. }\label{tab:sizevsrange}}
%\end{minipage}
%\begin{tabular}{ccccccc}
%\vspace*{-3mm}\\
%\noalign{\hrule height 0.8pt}
%$\langle r^2\rangle^{1/2}$(fm)& $R_S$(fm)& $m_S$(GeV)& $a_S$&
%$R_V$(fm)&$m_V$(GeV)& $a_V$\\
%\hline 
%1.396 & 2.260 &0.2140 & 6.117 & 1.640  &0.2947 &0.4092\\
%0.7708 & 1.128 &0.4283 & 14.55 & 0.8713 &0.5547 &1.742\\
%0.5381 &0.7208 &0.6706 & 27.44 & 0.5663 &0.8535 &5.129\\
%0.4896 &0.6310 &0.7661 & 33.20 & 0.4980 &0.9705 &7.185\\
%0.4586 &0.5710 &0.8464 & 38.28 & 0.4522 &1.069  &9.331\\
%0.4375 &0.5280 &0.9154 & 42.75 & 0.4191 &1.153  &11.55\\
%\noalign{\hrule height 0.8pt}
%\end{tabular}
%\end{center}
%}
%\end{table}
%%table-------------------------------------------------

%figure5-------form factors for various binding energy---------------
%figure---------------------------------------------

When the previous parameter set is used, the interaction mass 
parameter is about 550 MeV and 600 MeV for vector and scalar 
interactions, respectively.  
However, if we use a parameter set that produces 
the nucleon size about 
0.6 fm, then the two masses are 
$m_S \sim 600$ MeV and 
$m_V \sim 800$ MeV.  
These numbers are remarkably close to the mass of the sigma and omega
mesons.  
The nucleon size of 0.6 fm should be reasonable as the 
size of the nucleon core.  
Meson clouds around it contributes to a substantial 
amount to the size of the nucleon~\cite{Hosaka:1996ee}.  

%figure6---------------------------------------------
%figure---------------------------------------------

In the present analysis, apart from the absolute value of the 
interaction strengths, the interaction ranges for the scalar
and vector interactions have been produced appropriately by the 
quark exchange picture.  
One question concerns the small (but non-negligible) difference
between the ranges in the scalar and vector channels, which is
consistent with the empirical fact.  
It is crudely understood by the dimensionality of the loop 
integral.  
As seen from Eq.~(\ref{eqn:NNN}) 
the integrand for the vector interaction has higher 
power in momentum variable than the scalar one.  
Because of this, the vector part reflects  
shorter-distant dynamics and produces shorter interaction 
range.  

%==================================================================
\section{Summary}
%==================================================================

In this paper we have studied the nucleon-nucleon (NN) interaction 
using a microscopic theory of quarks and diquarks.  
Nucleons are described as quark-diquark bound states, 
which was shown to 
explain reasonably well their static 
properties~\cite{Abu-Raddad:2002pw}.  
%Furthermore, the model satisfies important symmetries such as the 
%gauge and chiral symmetries.  
The quark and diquark degrees of freedom were integrated out in the 
path-integral method and an effective lagrangian was derived for 
mesons and baryons.  
The resulting trace-log formula contains various meson and 
nucleon interaction terms, including the NN interaction 
in the short range region expressed by a Feynman integral 
of a quark-diquark loop.  
Hence the nucleon-nucleon interaction is naturally expressed as meson 
exchanges at long ranges, while quark and diquark 
exchanges at short ranges.  
The NN interaction contains the scalar and vector 
type terms at leading orders in powers of momentum of the 
Feynman integral.  
It turns out that the scalar term is attractive, while the vector 
term repulsive.  

In the present paper our numerical calculations 
contained only scalar 
diquarks as a first step toward full calculations.  
Therefore, the magnitude of the interaction strengths were 
not studied; they must be affected by the inclusion of 
the axial-vector diquark.  
On the other hand the interaction range was better studied
in the present framework, reflecting the size 
of the nucleon.
Consequently, the range of the scalar and vector interactions 
came out to be quite reasonable; in the mass parameter, 
the scalar mass was about 600 MeV and the vector mass about 
800 MeV, when the nucleon size was set at around 0.6 fm.  

The present result encourages us to study further baryon
properties by extending 
the model by including the axial-vector diquark.  
We have already started the study of such direction.  
When the axial vector is included, due to the massive vector nature of the 
propagator the loop integral diverges more strongly than the case of 
the scalar diquark.  
This complicates numerical study more than the present case.  
We hope to make progress along this line in the future.  

%By fine tuning the microscopic treatment for mesons and baryons 
%with their structure and interactions treated in a single framework 
%is extremely important when we consider the properties of 
%hadronic matter in extreme conditions at finite temperature, 
%density and perhaps with finite strangeness.  
%In such circumstances, the change in hadron structure is very 
%important with respecting symmetries of the system.  
%

\section*{Acknowledgements}

We would like to thank Laith Abu-Raddad and Dietmar Ebert for 
discussions and comments at the early stage of this work.
We would also like to thank Hiroshi Toki for valuable 
comments and encouragements. 

%
%  Figure 
% 


\begin{thebibliography}{99}

\bibitem{NJL}
Y. Nambu and G. Jona-Lasinio, Phys. Rev. {\bf 122} (1961) 345; 
{\it ibid.} {\bf 124} (1961) 246.

\bibitem{Abu-Raddad:2002pw}
L.~J.~Abu-Raddad, A.~Hosaka, D.~Ebert and H.~Toki,
Phys.\ Rev.\ C {\bf 66} (2002) 025206.  

%=========
%  NN int.
%=========

%\cite{Lacombe:dr}
\bibitem{Lacombe:dr}
M.~Lacombe, B.~Loiseau, J.~M.~Richard, R.~Vinh Mau, J.~Cote, P.~Pires and R.~De Tourreil,
%``Parametrization Of The Paris N N Potential,''
Phys.\ Rev.\ C {\bf 21} (1980) 861.
%%CITATION = PHRVA,C21,861;%%

%\cite{Machleidt:hj}
\bibitem{Machleidt:hj}
R.~Machleidt, K.~Holinde and C.~Elster,
%``The Bonn Meson Exchange Model For The Nucleon Nucleon Interaction,''
Phys.\ Rept.\  {\bf 149} (1987) 1.
%%CITATION = PRPLC,149,1;%%

%%\cite{Machleidt:1992uz}
%\bibitem{Machleidt:1992uz}
%R.~Machleidt and G.~Q.~Li,
%%``Nucleon-nucleon potentials in comparison: Physics or polemics?,''
%Phys.\ Rept.\  {\bf 242} (1994) 5
%[arXiv:nucl-th/9301019].
%%%CITATION = NUCL-TH 9301019;%%

%\cite{Toki:ai}
\bibitem{Toki:ai}
H.~Toki,
%``Short Range Behavior Of The N N Potential Within The Quark Model,''
Z.\ Phys.\ A {\bf 294} (1980) 173.
%%CITATION = ZEPYA,A294,173;%%

%\cite{Oka:rj}
\bibitem{Oka:rj}
M.~Oka and K.~Yazaki,
%``Short Range Part Of Baryon Baryon Interaction In A Quark Model. Ii. Numerical Results For S Wave,''
Prog.\ Theor.\ Phys.\  {\bf 66} (1981) 556; {\it ibid.} 572.
%%CITATION = PTPKA,66,572;%%
%%\cite{Oka:ri}
%\bibitem{Oka:ri}
%M.~Oka and K.~Yazaki,
%%``Short Range Part Of Baryon Baryon Interaction In A Quark Model. I. Formulation,''
%Prog.\ Theor.\ Phys.\  {\bf 66} (1981) 556.
%%%CITATION = PTPKA,66,556;%%

%\cite{Takeuchi:yz}
\bibitem{Takeuchi:yz}
S.~Takeuchi, K.~Shimizu and K.~Yazaki,
%``Nucleon-Nucleon Interaction In The Quark Cluster Model,''
Nucl.\ Phys.\ A {\bf 504} (1989) 777.
%%CITATION = NUPHA,A504,777;%%

%%\cite{Shimizu:wm}
%\bibitem{Shimizu:wm}
%K.~Shimizu, S.~Takeuchi and A.~J.~Buchmann,
%%``Study Of Nucleon Nucleon And Hyperon Nucleon Interaction,''
%Prog.\ Theor.\ Phys.\ Suppl.\  {\bf 137} (2000) 43.
%%%CITATION = PTPSA,137,43;%%

%\cite{Fujiwara:2001pw}
\bibitem{Fujiwara:2001pw}
Y.~Fujiwara, T.~Fujita, M.~Kohno, C.~Nakamoto and Y.~Suzuki,
%``Resonating-group study of baryon-baryon interactions for the complete baryon octet: NN interaction,''
Phys.\ Rev.\ C {\bf 65} (2001) 014002.
%%CITATION = PHRVA,C65,014002;%%

%%\cite{Fujiwara:sv}
%\bibitem{Fujiwara:sv}
%Y.~Fujiwara, T.~Fujita, C.~Nakamoto, Y.~Suzuki and M.~Kohno,
%%``Baryon Baryon Interaction In A Quark Model,''
%Nucl.\ Phys.\ A {\bf 639} (1998) 41.
%%%CITATION = NUPHA,A639,41;%%



%\cite{Eguchi:1976iz}
\bibitem{Eguchi:1976iz}
T.~Eguchi,
%``A New Approach To Collective Phenomena In Superconductivity Models,''
Phys.\ Rev.\ D {\bf 14} (1976) 2755.
%%CITATION = PHRVA,D14,2755;%%

%\cite{Dhar:1983fr}
\bibitem{Dhar:1983fr}
A.~Dhar and S.~R.~Wadia,
%``The Nambu-Jona-Lasinio Model: An Effective Lagrangian For Quantum Chromodynamics At Intermediate Length Scales,''
Phys.\ Rev.\ Lett.\  {\bf 52} (1984) 959.
%%CITATION = PRLTA,52,959;%%

%\cite{Ebert:1985kz}
\bibitem{Ebert:1985kz}
D.~Ebert and H.~Reinhardt,
%``Effective Chiral Hadron Lagrangian With Anomalies And Skyrme Terms From Quark Flavor Dynamics,''
Nucl.\ Phys.\ B {\bf 271} (1986) 188.
%%CITATION = NUPHA,B271,188;%%

%\cite{Ebert:1997hr}
\bibitem{Ebert:1997hr}
D.~Ebert and T.~Jurke,
%``Effective chiral meson baryon Lagrangian from path integral  hadronization of a quark diquark model,''
Phys.\ Rev.\ D {\bf 58} (1998) 034001. 
%[arXiv:hep-ph/9710390].
%%CITATION = HEP-PH 9710390;%%

%\cite{Ishii:2000zy}
\bibitem{Ishii:2000zy}
N.~Ishii,
%``Goldberger-Treiman relation and g(pi N N) from the three quark  BS/Faddeev approach in the NJL model,''
Nucl.\ Phys.\ A {\bf 689} (2001) 793
%[arXiv:nucl-th/0004063].
%%CITATION = NUCL-TH 0004063;%%

\bibitem{espriu}
D. Espriu, P. Pascual, and R. Tarrach, Nucl. Phys. B214, 285 (1983).

\bibitem{weinberg_book}
S. Weinberg, {\it The Quantum Theory of the Fields}, 
Cambridge (1995).  

\bibitem{hosaka_book}
A. Hosaka and H. Toki, 
{\it Quarks, Baryons and Chiral Symmetry}, World Scientific (2001).   

\bibitem{Goldberger:1958tr}
M. L. Goldberger and S. B. Treiman, 
Phys. Rev. {\bf 110} (1958) 1178.  

\bibitem{izykson}
C.~Izykuson and J.B.~Zuber, {\it Quantum Field Theory} McGraw-Hill (1980).

\bibitem{Machleidt:1987hj}
R. Machleidt and K. Holinde and C. Elster, 
Phys. Rept. {\bf 149} (1987) 1.  

%\cite{Liesenfeld:1999mv}
\bibitem{Liesenfeld:1999mv}
A.~Liesenfeld {\it et al.}  [A1 Collaboration],
%``A measurement of the axial form-factor of the nucleon by the  p(e,e' pi+)n reaction at W = 1125-MeV,''
Phys.\ Lett.\ B {\bf 468} (1999) 20.  
%[arXiv:nucl-ex/9911003].
%%CITATION = NUCL-EX 9911003;%%


\bibitem{Hosaka:1996ee}
A. Hosaka and H. Toki, Phys. Rept. {\bf 277} (1996) 65.  

\end{thebibliography}
\end{document}